\documentclass{iopconfser/iopconfser}

\usepackage{amsmath,amssymb} %math definitions
\usepackage{graphicx} %\includegraphics
\usepackage{subfig} %\subfloat
\usepackage[export]{adjustbox} % valign=c for \subfloat and \includegraphics
\usepackage{appendix} %\appendix
\usepackage{xfrac} %\sfrac
\usepackage{float} %[H] for placing figures, tables etc.
\usepackage{courier} %\texttt

\usepackage[numbers,square,comma,sort]{natbib} 

\usepackage{csquotes} %\enquote

\usepackage[colorlinks=true,linkcolor=blue,
            urlcolor=blue,citecolor=red,
            pdfusetitle]{hyperref} %customize link colors
\usepackage[nameinlink,capitalize]{cleveref} %\cref for automatic Eq(s). 1 formatting

\graphicspath{
    {./Figures}
    {./Figures/PenroseDiagramInUpMode}
    {./Figures/BHWaveShadowSchemes}
    {./Figures/Manko-Ruiz}
    {./Figures/ObserverSourceLocation}
    {./Figures/Results}
}

\begin{document}
\title{Wave optical imaging by point-source scattering for a TNdS black hole}
\author{Felix Willenborg$^{1, 2}$}
\affil{$^1$Zentrum für angewandte Raumfahrttechnologie und Mikrogravitation (ZARM), \\
Unversity of Bremen, 28359 Bremen, Germany \vspace{3mm}
}
\affil{$^2$
Gauss-Olbers Space Technology Transfer Center, \\
c/o ZARM, Unversity of Bremen, 28359 Bremen, Germany
}
\email{felix.willenborg@zarm.uni-bremen.de}

\begin{abstract}
In our work, we present a calculation based on an exact, non-approximated wave equation, as described by the Teukolsky equation, for a Taub-NUT-de Sitter spacetime. 
We observe the scattering of a monochromatic point source by an observer located at a greater distance from the black hole, and examine the wave-optical images by a single source and by multiple sources. 
\end{abstract}

\section{Introduction}
Wave-optical imaging provides an alternative approach to gravitational lensing by compact objects.
Rather than relying on approximations via the so-called amplification factor $F$ \cite{Schneider1999} (e.g. in Ref. \cite{Cheung2021}) or ray-optical approaches \cite{Grenzebach2015} (e.g. close to our approach in Ref. \cite{Frost2022}), wave-optical imaging uses the full exact wave equation. 
In Ref. \cite{Nambu2012}, wave-optical imaging was first introduced by a \enquote{telescope with a single convex thin lens}, and in Ref. \cite{Motohashi2021} the solvability of the Kerr-de Sitter Teukolsky equations was demonstrated. 
Our previous work \cite{Willenborg2018} built on these to achieve the desired full exact wave-optical imaging.
Here, we want to extend to the Taub-NUT-de Sitter spacetime in order to further examine the wave-optical imaging, compare our approach with ray-optical results and finally draw conclusions about the wave-optical shadow.

\section{Taub-NUT-de Sitter Teukolsky master equation}
The Teukolsky master equation is a second-order partial differential equation that describes the first-order perturbations of the background spacetime. 
It is derived from the Taub-NUT-de Sitter spacetime via a reformulation in the Newman-Penrose formalism (see. Refs. \cite{Teukolsky_1973,Newman1962}). 
Due to the fact that the differential equation is separable (see Ref. \cite{Kamran1987}), using the Ansatz
\begin{align}
\Psi_{nm}(t,r,\phi,\theta) = R_{nm}(r) ~S_{nm}(\theta) e^{-i \omega t} e^{i m \phi},
\end{align}
where $n$ is the node index and $m$ the azimuthal index, two separated and ordinary second-order differential equations are received:
\begin{align}
\frac{d}{dr}\left(\Delta_r \frac{d}{dr} R_{nm}(r) \right) + V^\text{(rad)}_{nm}(r) \,R_{nm}(r) &= 0 \label{TME:Eq:Rad}
\end{align}
and 
\begin{align}
\frac{d}{dx} \left((1 - x^2) \frac{d}{dx} S_{nm}(x)\right) + V^\text{(ang)}_{nm}(x) \,S_{nm}(x) &= 0, \label{TME:Eq:Ang}
\end{align}
where $x = \cos\theta$.
$V^\text{(rad)}_{nm}(r)$ and $V^\text{(ang)}_{nm}(x)$ are functions not written out in full for the sake of the overview. 
However, they satisfy the condition that transforms \cref{TME:Eq:Rad,TME:Eq:Ang} into Sturm-Liouville type differential equations, where the eigenvalue $\lambda_{nm}$ is contained within $V^\text{(rad)}_{nm}(r)$ and $V^\text{(ang)}_{nm}(x)$.
Since the Teukolsky equations are \enquote{Heun equations in disguise} \cite{Batic_2007}, a transformation into the (confluent) Heun equation will be the next step to solve them.

\section{Solution of separated Teukolsky equation}
The (confluent) Heun equation is the most general second-order ordinary differential equation with (two) four regular singularities \cite{Ronveaux1995}.
Its solutions are computationally implemented in various projects (see Refs. \cite{Motygin2015,Birkandan2021a}). 
We rely on the Wolfram Mathematica 12.3 implementations in the following (cf. \cite{WolframMathematicaHeunC2020,WolframMathematicaHeunG2020}). 

\subsection{Angular Teukolsky equation}
\cref{TME:Eq:Ang} can be transformed into the confluent Heun equation, since it is a second-order ordinary differential equation with two regular singularities at $x \in \{-1, 1\}$, and an irregular singularity at $x = \infty$.
Appropriate transformations of the independent and dependent variables convert \cref{TME:Eq:Ang} into the confluent Heun equation \cite{Ronveaux1995}
\begin{align}
    y''(z) + \left(\frac{\gamma}{z} + \frac{\delta}{z - 1} + \epsilon\right) y'(z) + \left(\frac{\alpha z - \sigma}{z (z - 1)}\right) y(z) = 0 \label{TMESol:Ang:Eq:HeunC} \, .
\end{align}
The function $y(z)$, which describes a so-called confluent local Heun function $CHl$, is regular at both poles/regular singularities, i.e. at $z \in \{0, 1\}$. 
This can be achieved using a very specific form of the eigenvalue $\lambda_{nm}$, which transforms the $CHl$ into the confluent Heun function $CHf$ (cf. \cite{Willenborg2024a} for the Taub-NUT case). 
This establishes an orthogonality relationship between the possible solutions. 
Thus, a complete set of orthonormal functions $\tilde S_{nm}(x)$ can be constructed, giving the completeness relation
\begin{align}
    \sum\limits^\infty_{n = 0} \sum\limits^{\infty}_{m = -\infty} \tilde S_{nm}(x) \tilde S_{nm}(x') = \delta(x - x') \, . \label{TMESol:Ang:Eq:CompletenessRelMisner}
\end{align}

\subsection{Radial Teukolsky equation}
We repeat the initial steps of the angular Teukolsky equation for the radial Teukolsky equation. 
This time, four regular singularities have to be covered.
These steps turn \cref{TME:Eq:Rad} into
\begin{align}
y''(z) + \left(\frac{\gamma}{z} + \frac{\delta}{z - 1} + \frac{\epsilon}{z - a_H}\right) y'(z)  + \frac{\alpha \beta z - q}{z (z - 1) (z - a_H)} y(z) = 0 \, ,
\label{Heun:Eq:DEQ}
\end{align}
the general Heun equation \cite{Ronveaux1995} with $y(z)$ describing the local Heun functions $Hl$.
Two linearly independent solutions, $R_\text{in}$ and $R_\text{up}$, are constructed using the boundary conditions that specify either ingoing modes at the event horizon $\mathcal{H}_0$ or outgoing modes at the cosmological horizon $\mathcal{H}_c$. 

\subsection{Point-source scattering of scalar waves}
\begin{figure}[H]
\centering
\subfloat[]{\includegraphics[height=0.28\textwidth,valign=c]{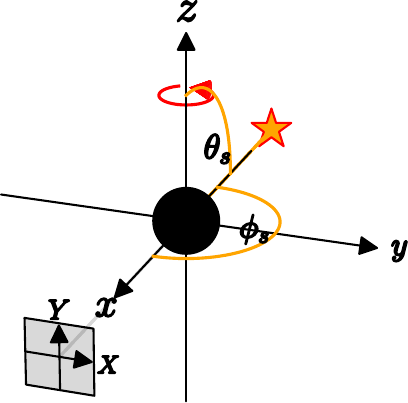}\label{TMESol:Scatter:Fig:Schemes:Single}}
\subfloat[]{\includegraphics[width=0.28\textwidth,valign=c]{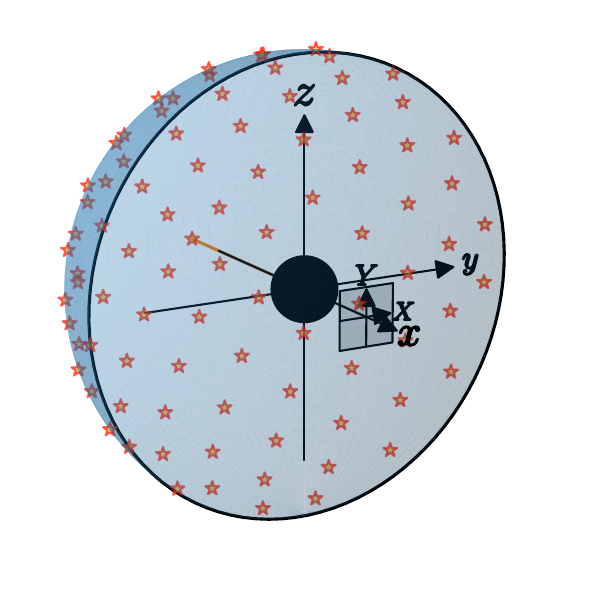}\label{TMESol:Scatter:Fig:Schemes:Mult}} 
\caption{\textbf{a)} Coordinate observer plane-point source alignment, \textbf{b)} Regularly distributed sources by Lebedev quadrature on an half sphere}
\label{TMESol:Scattering:Fig:Schemes}
\end{figure}

Our aim is to examine the scattering of a point-source with a monochromatic frequency of $\omega$ by a Taub-NUT-de Sitter black hole. 
To achieve this, we introduce a point-source term into the Teukolsky master equation at $\{r_S = 20 M, \theta_S = \frac{\pi}{2}, \phi_S = \pi\}$. 
Ultimately, we arrive at
\begin{align}
G(\vec{r}, \vec{r}_s, L) = \sum\limits^{L}_{n = 0} \sum\limits^\infty_{m = -\infty} &\tilde{G}_{nm}(r,r_s) \tilde S_{nm}(\theta) \tilde S_{nm}^*(\theta_s) e^{i m \phi} e^{-i m \phi_s} \, . \label{TMESol:Scatter:Eq:GreensFunc}
\end{align}
The series is stopped at $L$ to heuristically ensure  convergence of the series (cf. Ref. \cite{Willenborg2024}). 
The radial solution $\tilde{G}_{nm}(r,r_s)$ combines $R_\text{in}$ and $R_\text{up}$, such that they coincide at $r = r_s$.
For the purposes of observation, we construct a square coordinate observer plane, with its center at $\{r_O = 10M, \theta_O = \frac{\pi}{2}, \phi_O = 0\}$ (cf. the gray square in \cref{TMESol:Scatter:Fig:Schemes:Single}). 
The length of each edge of the coordinate plane will be set to $d = 2 M$. 
The scattering of the point-source waves by the Taub-NUT-de Sitter black hole will be evaluated using \cref{TMESol:Scatter:Eq:GreensFunc} on a $71 \times 71$ grid.
Since we are examining scalar waves, we can make use of the Kirchhoff-Fresnel approximation in order to calculate the wave-optical image from the evaluation on the coordinate plane, see Refs. \cite{Willenborg2024,Nambu2012,Nambu2016} for more details.

\section{Results of point-source scattering}
\begin{figure}[H]
\centering
\subfloat[$N = 0.00M$ (SdS)]{\includegraphics[width=0.28\linewidth,valign=c]{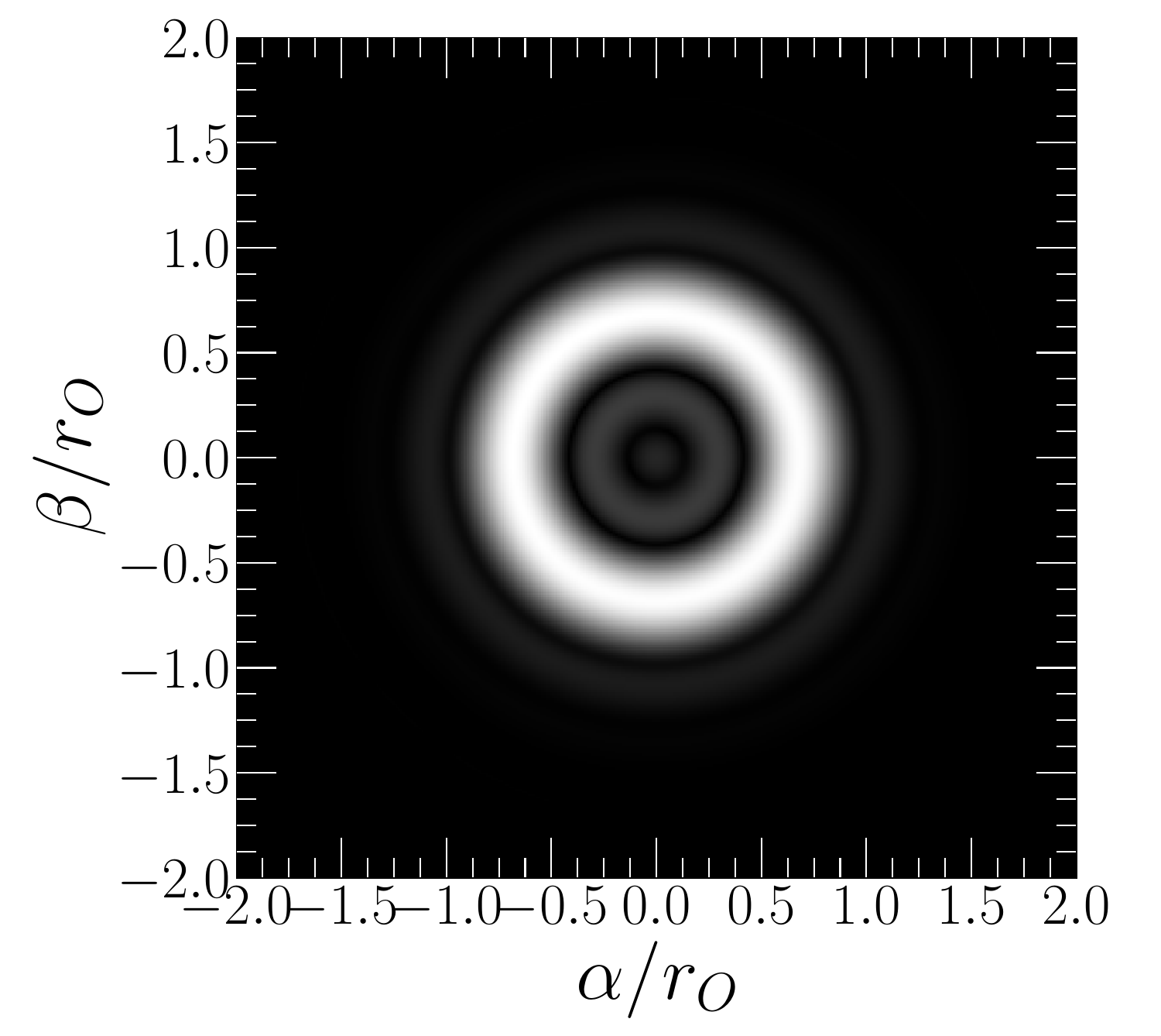}\label{Result:Fig:Result:N0FFT}}
\subfloat[$N = 0.10M$]{\includegraphics[width=0.28\linewidth,valign=c]{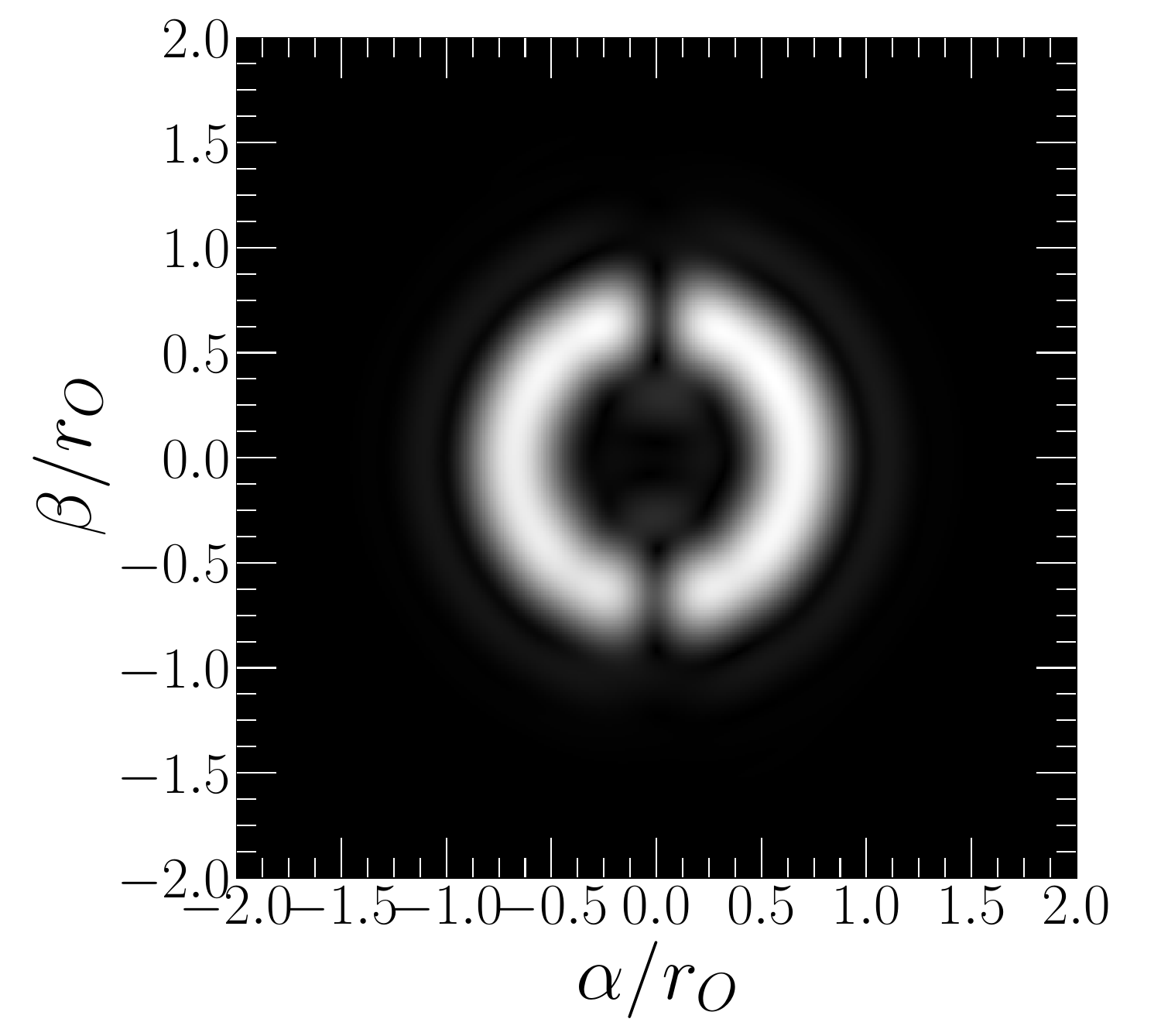}\label{Result:Fig:Result:N0.10FFT}}
\subfloat[$N = 0.22M$]{\includegraphics[width=0.28\linewidth,valign=c]{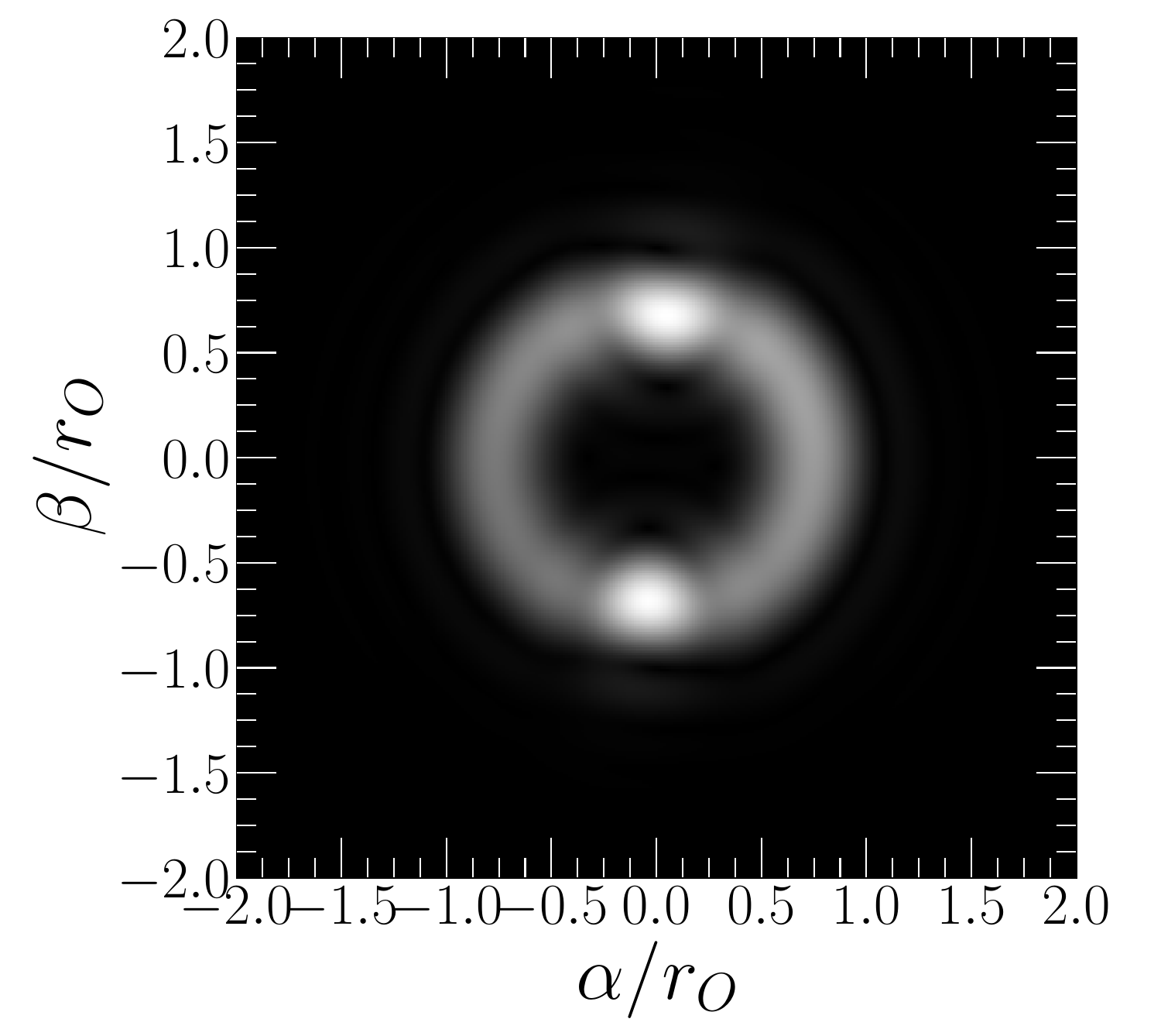}\label{Result:Fig:Result:N0.22FFT}}  \\
\subfloat[$N = 0.00M$ (SdS)]{\includegraphics[width=0.28\linewidth]{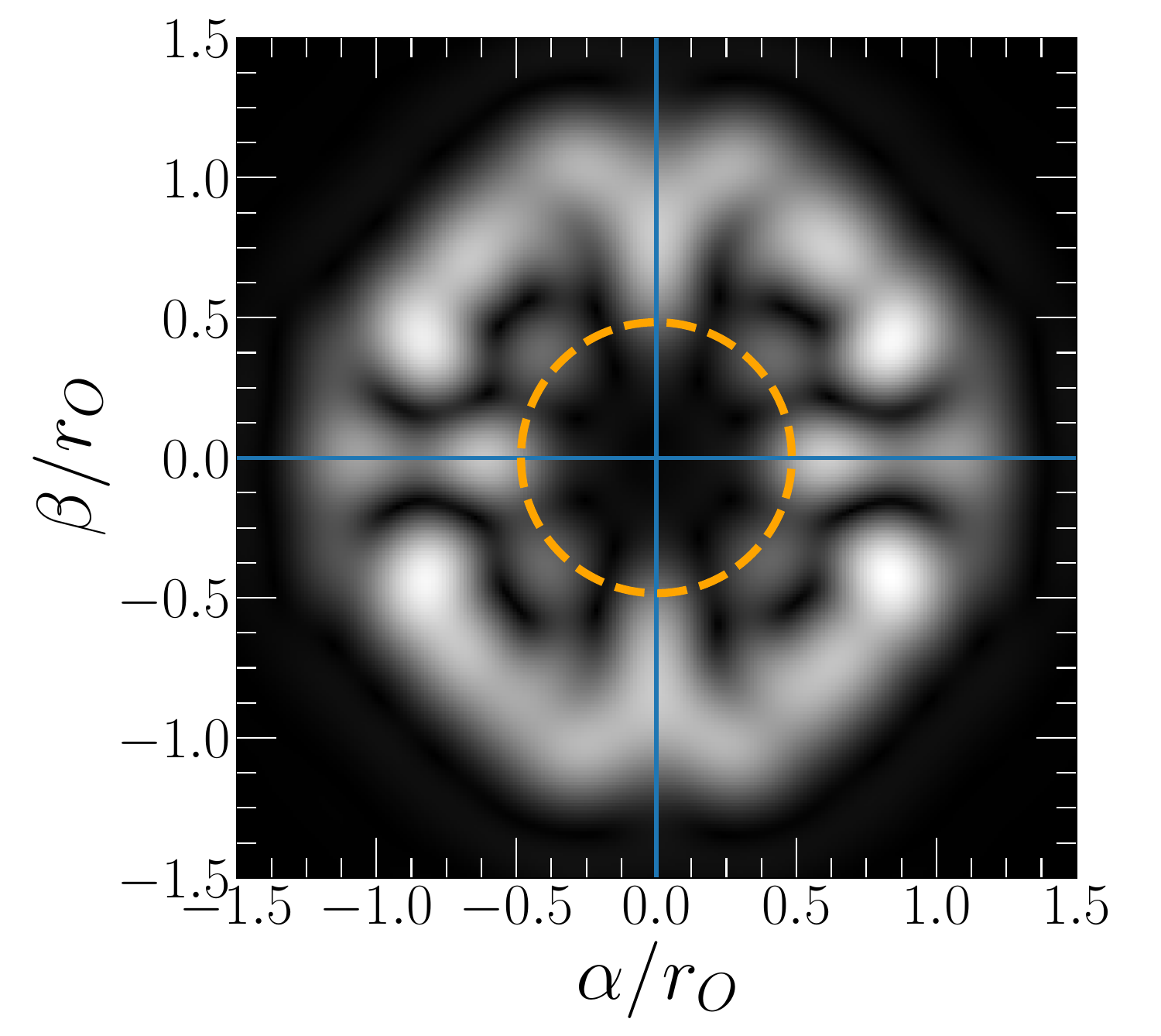}\label{Result:Fig:Result:N0FFTSh}}
\subfloat[$N = 0.25M$]{\includegraphics[width=0.28\linewidth]{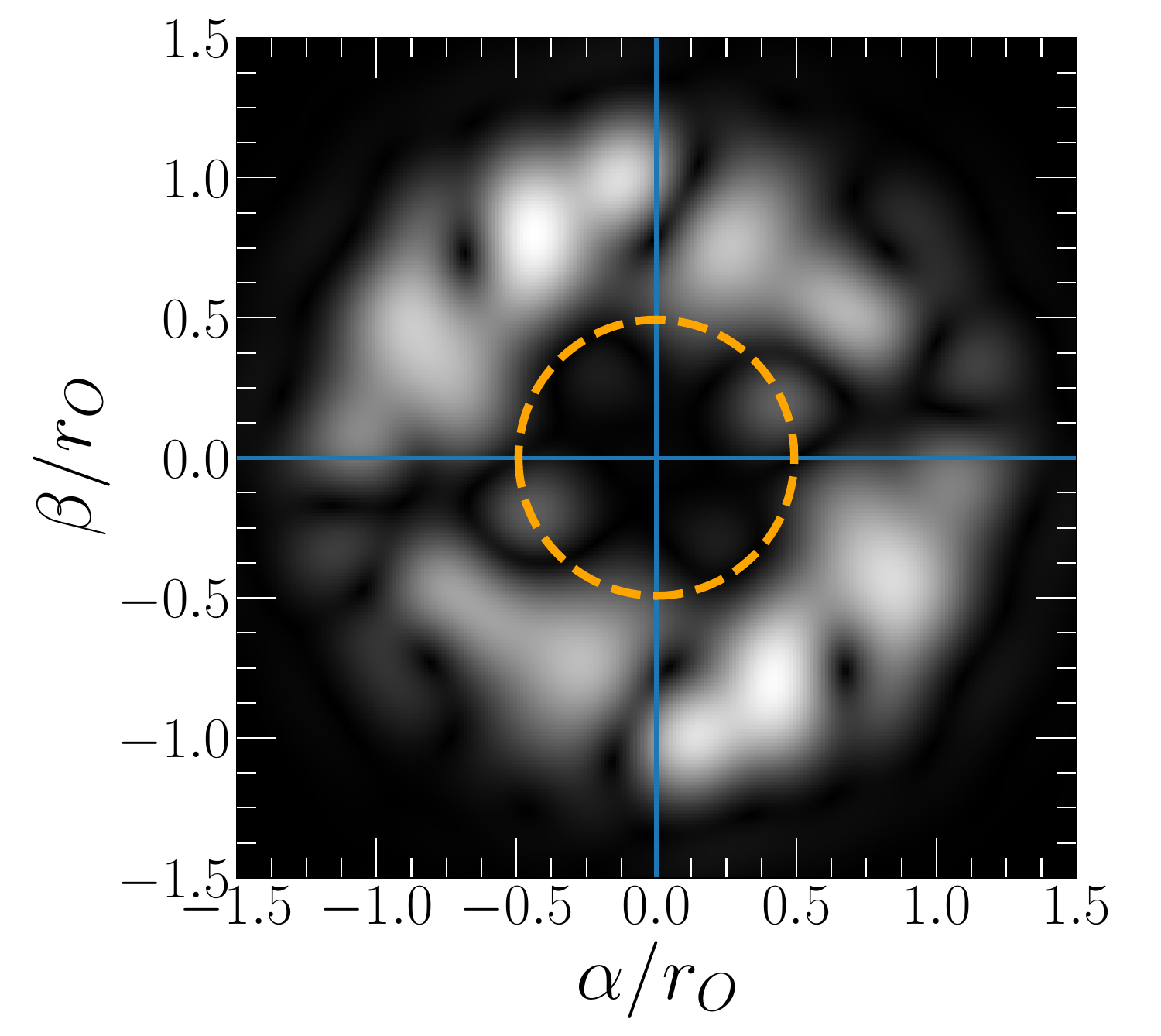}\label{Result:Fig:Result:N0.25FFTSh}}
\subfloat[$N = 0.50M$]{\includegraphics[width=0.28\linewidth]{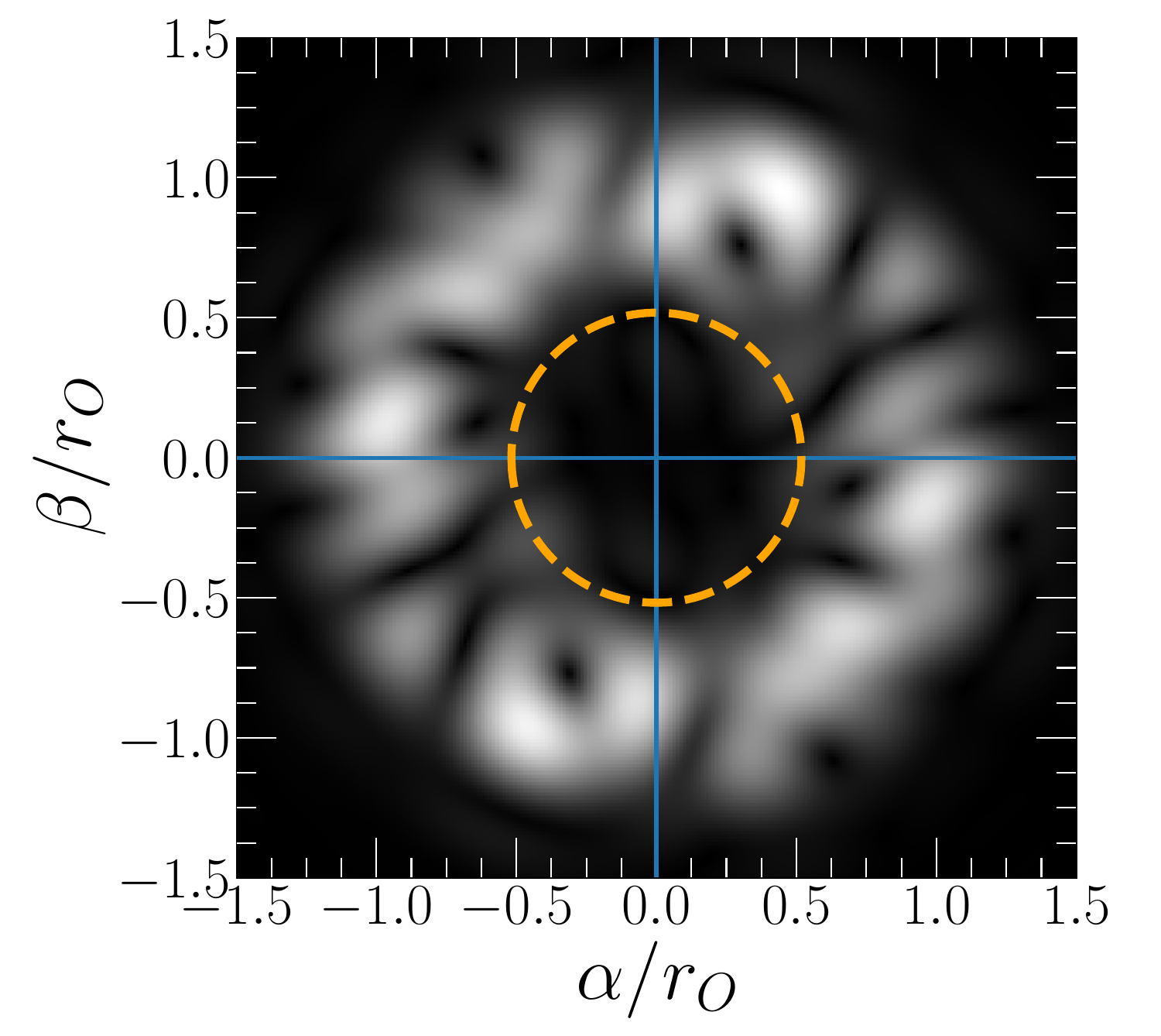}\label{Result:Fig:Result:N0.50FFTSh}} 
\caption{Variation of $N$ for fixed $M\omega = 8$, $C = 0$. \textbf{a)-c)}: Single point-source, \textbf{d)-f)}: multiple point-sources, see \cref{TMESol:Scatter:Fig:Schemes:Mult}. Orange-dashed lines are ray-optical shadow predictions by Ref. \cite{Grenzebach2016}.}
\label{Result:Fig:Result}
\end{figure}

Throughout the following, we will use the following frequency $M\omega = 8$, the cosmological constant $\Lambda M^2 = 10^{-3}$, and the Manko-Ruiz parameter $C = 0$. 
The Manko-Ruiz parameter, introduced in Ref. \cite{Manko2005}, controls the semi-infinite singularities of the Taub-NUT spacetime. 
For $C = 0$, both axes have semi-infinite singularities that induce the same infinite amount of total angular momentum to the spacetime. 
However, they are counter-rotating. 
Thus, the total angular momentum induced by the semi-infinite singularities vanishes.
The resulting wave-optical image for a single point-source, for the configuration shown in \cref{TMESol:Scatter:Fig:Schemes:Single}, is given in \cref{Result:Fig:Result:N0FFT,Result:Fig:Result:N0.10FFT,Result:Fig:Result:N0.22FFT}.
\cref{Result:Fig:Result:N0FFT} shows the Schwarzschild-de Sitter case, for which an Einstein ring is predicted based on the chosen parameters. 
The wave-optical image shows this circumstance, but in a blurred way because of the lower frequency and its wave-optical nature.
In \cref{Result:Fig:Result:N0.10FFT} the NUT charge $N$ is the only parameter increased, resulting in a destructive interference at the apparent location of the semi-infinite singularities for the observer. 
Increasing $N$ furthermore, \cref{Result:Fig:Result:N0.22FFT} shows that also a constructive behavior is possible. 
Our results are consistent with the ray-optical results reported in Ref. \cite{Frost2022}, with the exception of the constructive and destructive behavior of the semi-infinite singularities. 
The configuration of multiple point-sources shown in \cref{TMESol:Scatter:Fig:Schemes:Mult} can be used to reveal how a wave-optical shadow might look like. 
This is because we construct a background whose radiation to the observer can sufficiently cover the photon-region. 
As expected, the Schwarzschild-de Sitter case in \cref{Result:Fig:Result:N0FFTSh} reveals the circular shape of the wave-optical shadow. 
Increasing the NUT charge $N$ causes the circular shadow to take on a more elliptical form, seen in \cref{Result:Fig:Result:N0.25FFTSh,Result:Fig:Result:N0.50FFTSh}.
This is not predicted by the ray-optical results and only applies for lower frequencies. 
Increasing the frequency $\omega$ for \cref{Result:Fig:Result:N0.25FFTSh,Result:Fig:Result:N0.50FFTSh}, turns the wave-optical shadows circular again -- as to be expected in order to align with ray-optical results. 

\section{Conclusions and outlook}
Wave-optical imaging reveals a destructive and constructive behavior in the apparent observations of semi-infinite singularities for single point-sources. 
In ray-optical results, this behavior must be assumed, whereas in our approach, it is simply a consequence of the wave-optical approach to gravitational lensing. 
In the case of multiple sources, the wave-optical shadow can be observed. 
It is shown that, for a certain set of parameters, the wave-optical shadow can become elliptical, which is not predicted by the ray-optical approach.
However, this may be an artifact of the multiple point-sources setup. 
Using a plane wave or extended source approach in  \cref{TMESol:Scatter:Eq:GreensFunc} might eliminate this behavior.
A full publication of this work is in preparation. 

\clearpage
\section{Acknowledgments}
The author would like to express gratitude to fruitful discussions with Claus Lämmerzahl, Dennis Philipp, Volker Perlick, Domenico Giullini and Torben Frost.
Additionally, gratitude is expressed for the authors of the software package {\texttt{xAct}} \cite{xAct} for Mathematica, which was used in the calculation of the Teukolsky master equation 
The research is funded by the Deutsche Forschungsgemeinschaft (DFG, German Research Foundation) – Project-ID 434617780 – SFB 1464 and
funded by the Deutsche Forschungsgemeinschaft (DFG, German Research Foundation) under Germany’s Excellence Strategy – EXC-2123 QuantumFrontiers – 390837967, and through the Research Training Group 1620 \enquote{Models of Gravity}.

%----------------------------------------------------------------------------------------
%	ZITATE/QUELLEN/LITHOGRAFIEN/BIBLIOGRAPHIE
%----------------------------------------------------------------------------------------
\providecommand{\newblock}{}

%----------------------------------------------------------------------------------------
%	DOKUMENTENENDE
%----------------------------------------------------------------------------------------
\end{document}